\documentclass[12pt]{article} 
\usepackage{amsmath}
\usepackage{amssymb}
\usepackage{amstext}
\usepackage{bm}
\usepackage{graphicx}
\usepackage{epstopdf}
\usepackage{cite}

 \newcommand{\bl}{\big<}
  \newcommand{\bg}{\big>}

\allowdisplaybreaks 

\begin{document}
\baselineskip= 0.225in
\parindent=0.35in

%%%%%%%%%%%%%%%%%%%%%%%%%%%%%%%%%%%%%%%%%%%%%%%%%%

\vspace*{0.2in}
\begin{center} 
{\bf \Large Non-classical transport with\\ 
angular-dependent path-length distributions.\\
1: Theory\\ }
\vspace {0.5in} 
Richard Vasques \\
RWTH Aachen University, Aachen, Germany \\ 
Department of Mathematics\\
Center for Computational Engineering Science\\
vasques@mathcces.rwth-aachen.de \\
\vspace {0.2in} 
and \\
\vspace {0.2in} 
Edward W.\ Larsen \\ 
University of Michigan, Ann Arbor, U.S.A. \\
Department of Nuclear Engineering and Radiological Sciences \\ 
edlarsen@umich.edu \\
\end{center}

%%%%%%%%%%%%%%%%%%%%%%%%%%%%%%%%%%%%%%%%%%%%%%%%%

\vspace{5pt}
\begin{center}
{\bf ABSTRACT}
\end{center}
\begin{quote}
\begin{small}
\vspace{5pt}

This paper extends a recently introduced theory describing particle transport for random statistically homogeneous systems in which the distribution function $p(s)$ for chord lengths between scattering centers is non-exponential. Here, we relax the previous assumption that $p(s)$ does not depend on the direction of flight $\bm\Omega$; this leads to a new generalized linear Boltzmann equation that includes angular-dependent cross sections, and to a new generalized diffusion equation that accounts for anisotropic behavior resulting from the statistics of the system.

\end{small} 
\end{quote}

\setlength{\baselineskip}{0.225in}

%%%%%%%%%%%%%%%%%%%%%%%%%%%%%%%%%%%%%%%%%%%%%%%%%

\vspace{10pt}
\noindent
{\bf 1. INTRODUCTION}
\setcounter{section}{1}
\setcounter{equation}{0} 
\vspace{10pt}

The classical theory of linear particle transport defines as $dp = \Sigma_t(\bm x,E)ds$ the incremental probability $dp$ that a particle at point $\bm x$ with energy $E$ will experience an interaction while traveling an incremental distance $ds$ in the background material. Here, the total cross section $\Sigma_t$ is independent of the direction of flight $\bm\Omega$, and the path-length $s$ is defined as 
\begin{equation}
s =  \begin{array}{l}
\text{ the path-length traveled by the particle since }\\%\vspace{-0.4cm}\\
\text{ its previous interaction  (birth or scattering) . }
\end{array}
\label{1.1}
 \end{equation}
This typically leads to the particle flux decreasing as an exponential function of the path-length (Beer-Lambert law).

However, in an inhomogeneous random medium, particles will travel through different materials with randomly located interfaces. In atmospheric clouds, experimental studies have found evidence of a non-exponential attenuation law \cite{davis_96,marshak_97,pfeilsticker_99}. It has been suggested \cite{kostinski_01} that the locations of the scattering centers (in this case water droplets) are spatially correlated in ways that measurably affect radiative transfer within the cloud \cite{kostinski_01b,buldyrev_01, shaw_02,none_02,davis_08,borovoi_02,kostinski_02,davis_04,scholl_06}.

An approach to this type of non-classical transport problem was recently introduced \cite{larsen_07}, with the assumption that the positions of the scattering
 centers are correlated but independent of direction $\bm \Omega$; that is,
$\Sigma_t$ is independent of $\bm \Omega$ but not $s$: $\Sigma_t=\Sigma_t(\bm x, E, s)$. A full derivation of this generalized linear Boltzmann equation (GLBE) and its asymptotic diffusion limit can be found in \cite{larsen_11}, along with numerical results for an application in 2-D pebble bed reactor (PBR) cores. Existence and uniqueness of solutions, as well as their convergence to the diffusion equation, are rigorously discussed in \cite{frank_10}. Furthermore, a similar kinetic equation with path-length as an independent variable has been derived for the periodic Lorentz gas \cite{golse}.

For specific random systems in which the locations of the scattering centers are correlated and dependent on the direction $\bm\Omega$, anisotropic particle transport arises \cite{vasques_09,vasques_09b}. This anisotropy is a direct result of the geometry of the random system - for instance, the packing of pebbles close to the boundaries of a pebble bed system lead to particles traveling longer distances in directions parallel to the boundary wall \cite{vasques_13}. One may also expect that, due to the ``gravitational" arrangement of pebbles in PBR cores, diffusion in the vertical and horizontal directions might differ. This behavior can only be captured if we allow the path-lengths of the particles to depend upon $\bm\Omega$; that is, $\Sigma_t=\Sigma_t(\bm x, \bm\Omega, E, s)$. 

The goal of this paper is to extend the GLBE formulation in \cite{larsen_11} to include this angular dependence. For simplicity, we do not consider the most general problem here; similarly to \cite{larsen_11}, our analysis is based on five primary assumptions:
\begin{itemize}
\item[\textbf{i}] The physical system is infinite and statistically homogeneous.\vspace{-3pt}
\item[\textbf{ii}] Particle transport is monoenergetic. (However, the inclusion of energy- or frequency-dependence is straightforward.) \vspace{-3pt}
\item[\textbf{iii}] Particle transport is driven by a known interior isotropic source $Q(\bm x)$ satisfying $Q \to 0$ as $|\bm x| \to \infty$ (and the particle flux $\to 0$ as $|\bm x| \to \infty$). \vspace{-3pt}
\item[\textbf{iv}] The ensemble averaged total cross section $\Sigma_t(\bm\Omega,s)$, defined as
\begin{equation}
\Sigma_t(\bm\Omega,s)ds =  \begin{array}{l}
\text{ the probability (ensemble-averaged over all physical realiza-}\\
\text{ tions) that a particle, scattered or born at any point $\bm x$ and}\\
\text{ traveling in the direction $\bm\Omega$, will experience a collision}\\
\text{ between $\bm x + s\bm\Omega$ and $\bm x + (s+ds)\bm\Omega$,}\\
\end{array}\nonumber
 \end{equation}
 is known. (In the next part of this 2-part paper we discuss how $\Sigma_t(\bm\Omega,s)$ might be numerically derived from hypothesized correlations between the scattering centers.) \vspace{-7pt}
\item[\textbf{v}] The distribution function $P(\bm \Omega \cdot \bm \Omega')$ for scattering from $\bm \Omega'$ to $\bm \Omega$ is independent of $s$. (The correlation in the scattering center positions affects the probability of collision, but not the scattering properties when scattering events occur.)
\end{itemize}

For problems in general random media, $\Sigma_t(\bm\Omega,s)$ depends also on $\bm x$. In this paper the statistics are assumed to be homogeneous, in which case the dependence on $\bm x$ is dropped. (In the derivation of the GLBE in \cite{larsen_11}, the statistics was assumed to be independent of $\bm x$ and $\bm\Omega$.)

A summary of the remainder of the paper follows. In Section 2 we present definitions and formally derive the new GLBE. In Section 3 we derive (i) the conditional distribution function $q(\bm\Omega,s)$ for the distance $s$ to collision in a given direction $\bm\Omega$ in terms of the total cross section $\Sigma_t(\bm\Omega,s)$; and (ii) the equilibrium path-length spectrum in a given direction. In Section 4 we reformulate the new GLBE in terms of integral equations in which $s$ is absent. In Section 5 we derive the asymptotic diffusion limit of the new GLBE, presenting 3 physically relevant special cases; and in Section 6 we show that if $\Sigma_t(\bm\Omega,s)$ is independent of both $(\bm\Omega,s)$, the theory introduced here reduces to the classical theory. We conclude with a discussion in Section 7.

%%%%%%%%%%%%%%%%%%%%%%%%%%%%%%%%%%%%%%%%%%%%%%%%%
\vspace{20pt}
\noindent
{\bf 2. DERIVATION OF THE NEW GLBE}
\setcounter{section}{2}
\setcounter{equation}{0} 
\vspace{10pt}

    Using the notation $\bm x = (x,y,z) = $ position and $\bm \Omega = (\Omega_x, \Omega_y, \Omega_z) =$
direction of flight (with $|\bm \Omega | = 1$), and using Eq.\ (\ref{1.1}) for $s$, we define:
\begin{subequations} \label{2.1}
\begin{align}
\begin{array}{l}
     n(\bm x, \bm \Omega, s) dV d\Omega ds =
     \end{array} & \text{ the number of particles in } dV d\Omega ds
        \text{ about } (\bm x, \bm \Omega, s) ,
         \label{2.1a}\\
\begin{array}{l}     
     v =
     \end{array} & \; \frac{ds}{dt} = \text{ the particle speed , }   \\
     \begin{array}{l}
     \psi(\bm x, \bm \Omega, s) =
     \end{array} & \; v  n(\bm x, \bm \Omega, s) = \text{ the angular flux , } 
        \\
        \begin{array}{l}
        \Sigma_t(\bm\Omega, s) ds =\\
         \\
          \\
           \\
            \\
        \end{array}  &       
        \begin{array}{l}
        \text{the probability that a particle that has traveled}\\
        \text{a distance $s$ in the direction $\bm\Omega$ since its}\\
        \text{previous interaction (birth or scattering) will}\\
        \text{experience its next interaction while traveling a }\\
        \text{further distance $ds$,}
        \end{array}
        \\
        \begin{array}{l}
       c = \\
        \\
         \\
       \end{array} & \begin{array}{l}
       \text{the probability that when a particle experiences}\\
       \text{a collision, it will scatter (notice that $c$ is}\\
       \text{independent of $s$ and $\bm\Omega$),}
       \end{array}
          \\
          \begin{array}{l}
        P(\bm \Omega' \cdot \bm \Omega) d\Omega = \\
          \\
           \\
           \end{array} & \begin{array}{l}     
       \text{the probability that when a particle with direction}\\
       \text{of flight $\bm \Omega'$ scatters, its outgoing direction of flight}\\
       \text{will lie in $d \Omega$ about $\bm \Omega$ ($P$ is independent of $s$),}
       \end{array}
         \label{2.1f}
         \\ 
         \begin{array}{l}
         Q(\bm x) dV = \\
          \\
         \end{array} & \begin{array}{l}
       \text{the rate at which source particles are isotropically}\\%\vspace{-0.4cm}\\
       \text{emitted by an internal source $Q(\bm x)$ in $dV$ about $\bm x$.}
       \end{array} 
   \end{align}
   \end{subequations}
 Classic manipulations directly lead to:
   \begin{subequations} \label{2.2}
   \begin{align}
     \begin{array}{l} 
      \displaystyle{\frac{\partial}{\partial s} \psi(\bm x, \bm \Omega, s) dV d\Omega ds} =
      \end{array} & \frac{1}{v}\frac{\partial }{\partial t} 
         v  n(\bm x, \bm \Omega, s) dV d\Omega ds \\
      \begin{array}{l} = \\ \\ \end{array}  & \frac{\partial}{\partial t}  n(\bm x, \bm \Omega, s) dV d\Omega ds \nonumber \\
      \begin{array}{l} = \\ \\ \end{array} & \begin{array}{l}
          \text{the rate of change of the number of particles}\\%\vspace{-0.4cm}\\
          \text{in $dV d\Omega ds$ about $(\bm x, \bm \Omega, s)\, , $}
          \end{array}\nonumber
          \\  
          \begin{array}{l} 
      | \bm \Omega \cdot \textbf{n} | \psi (\bm x, \bm \Omega, s) dS d\Omega ds = \\ \\ \\
      \end{array} & \begin{array}{l}
       \text{the rate at which particles in $d\Omega ds$}\\%\vspace{-0.4cm} \\
       \text{about $(\bm \Omega,s)$ flow through an incremental}\\%\vspace{-0.4cm}\\
       \text{surface area $dS$ with unit normal vector \textbf{n},}
       \end{array} \\ 
       \begin{array}{l} 
      \bm \Omega \cdot \bm \nabla \psi (\bm x, \bm \Omega, s) dV d\Omega ds =\\ \\
      \end{array} & \begin{array}{l}
       \text{the net rate at 
         which particles in $d\Omega ds$}\\%\vspace{-0.4cm} \\
       \text{about $(\bm \Omega,s)$ flow (leak) out of $dV$ about $\bm x$,}
       \end{array} \\
       \begin{array}{l} 
      \Sigma_t(\bm\Omega, s) \psi(\bm x, \bm \Omega, s) dV d\Omega ds =
      \end{array} & \; \Sigma_t(\bm\Omega, s) \frac{ds}{dt} 
          n(\bm x, \bm \Omega, s) dV d\Omega ds \label{2.2d}\\
       \begin{array}{l} = \\ \end{array}  & \; \frac{1}{dt} [\Sigma_t(\bm\Omega,s) ds ] [ n(\bm x, \bm \Omega, s) dV d\Omega ds ] \nonumber \\
       \begin{array}{l} = \\ \\ \end{array}  & \begin{array}{l}
      \text{the rate at which particles in $dV d\Omega ds$ about}\\%\vspace{-0.4cm}\\
      \text{$(\bm x, \bm \Omega, s)$ experience collisions.}
      \end{array}\nonumber 
   \end{align}
The treatment of the inscattering and source terms requires extra care. From \linebreak
Eq.\ (\ref{2.2d}),
   \begin{align}
   \begin{array}{l}
     \displaystyle{ \left[ \int_0^{\infty} \Sigma_t(\bm\Omega', s') \psi(\bm x, \bm \Omega', s') ds'  \right] dV d\Omega'}
         = \\ \\
         \end{array}
         & \begin{array}{l}
         \text{the rate at which particles in $dV d \Omega'$}\\%\vspace{-0.4cm}\\
         \text{about $(\bm x, \bm \Omega')$ experience collisions.}
         \end{array}\nonumber
   \end{align}
Multiplying this expression by $c P(\bm \Omega \cdot \bm \Omega') d \Omega$, we obtain:   
   \begin{align}
     c P(\bm \Omega \cdot \bm \Omega') \left[ \int_0^{\infty} \Sigma_t(\bm\Omega', s')\psi(\bm x, \bm \Omega', s') ds'
          \right] dV d \Omega' d \Omega = &\nonumber \\
&\hspace{-5.0 cm} \begin{array}{l} = \\ \\ \end{array} \begin{array}{l}
\text{the rate at which particles in $dV d \Omega'$ about}\\%\vspace{-0.4cm}\\
\text{$(\bm x, \bm \Omega')$ scatter into $dV d\Omega$ about $(\bm x, \bm \Omega)$.}
\end{array}  \nonumber
   \end{align}
Integrating this expression over $\bm \Omega' \in 4\pi$, we get:
\pagebreak
   \begin{align}
      \left[ c \int_{4\pi} \int_0^{\infty} P(\bm \Omega' \cdot \bm
         \Omega) \Sigma_t(\bm\Omega', s') \psi(\bm x, \bm \Omega',s') \, ds' d
         \Omega' \right] dV d\Omega  \nonumber \hspace{1.4in} \\
      = \text{the rate at which particles scatter into $dV d\Omega$
        about $(\bm x, \bm \Omega) $ \;.}\nonumber 
   \end{align} 
Finally, when particles emerge from a scattering event their value of
$s$ is ``reset" to $s=0$. Therefore, the path-length spectrum of
particles that emerge from scattering events is the delta function
$\delta(s)$. Multiplying the previous expression by $\delta(s) ds$, we
obtain:
   \begin{align}
        \bigg[\delta(s) c \int_{4\pi} \int_0^{\infty} &P(\bm \Omega'
         \cdot
         \bm \Omega) \Sigma_t(\bm\Omega', s') \psi(\bm x, \bm \Omega',s') \, ds' 
         d \Omega' \bigg] dV d\Omega ds \\
      = &\text{ the rate at which particles scatter into $dV d\Omega ds$
        about $(\bm x, \bm \Omega,s) $ \;. }\nonumber
   \end{align} 
Also,
   \begin{align}
   \begin{array}{l}
     \displaystyle{ \delta(s) \frac{Q(\bm x)}{4\pi} dV d\Omega ds} = \\ \\
      \end{array} \begin{array}{l}
      \text{the rate at which source particles are}\\%\vspace{-0.4cm}\\
      \text{emitted into $dV d\Omega ds$ about $(\bm x, \bm
         \Omega,s)$.}
         \end{array}
   \end{align}
   \end{subequations}
 
 We now use the familiar conservation equation (in each of the
following terms, the phrase ``of particles in $dV d\Omega ds$ about 
$(\bm x, \bm \Omega, s)$" is omitted):
   \begin{align}
      \begin{array}{l} \text{Rate of change} = \end{array} & \; \text{Rate of gain} - \text{Rate of loss}
         \\
      \begin{array}{l} = \\ \\ \end{array} & \begin{array}{l}
      (\text{Inscatter rate} + \text{Source rate})\\%\vspace{-0.3cm} \\
       - (\text{Net leakage rate} + \text{Collision rate}) .
       \end{array}\nonumber
   \end{align}
Introducing Eqs.\ (\ref{2.2}) into this expression and dividing by $dV d\Omega ds $, we obtain the new GLBE for $\psi (\bm x, \bm \Omega, s)$:
   \begin{align}
      \frac{\partial \psi}{\partial s}&(\bm x, \bm \Omega, s)
          + \bm \Omega \cdot \bm \nabla \psi (\bm x, \bm \Omega, s)
         + \Sigma_t(\bm\Omega,s) \psi (\bm x, \bm \Omega, s)\label{2.4} \\
      & = \delta(s) \, c \int_{4\pi} \int_0^{\infty} P(\bm \Omega' \cdot
         \bm \Omega) \Sigma_t(\bm\Omega', s') \psi(\bm x, \bm \Omega',s') \, ds' 
         d \Omega' + \delta(s) \frac{Q(\bm x)}{4\pi} \,. \nonumber
   \end{align}

 Equation (\ref{2.4}) can be written in a mathematically equivalent way in which the delta function is not present: for $s>0$
   \begin{subequations} \label{2.5}
   \begin{equation}
      \frac{\partial \psi}{\partial s}(\bm x, \bm \Omega, s)
         + \bm \Omega \cdot \bm \nabla \psi (\bm x, \bm \Omega, s)
         + \Sigma_t(\bm\Omega, s) \psi (\bm x, \bm \Omega, s) = 0 \,;
   \label{2.5a}
   \end{equation}
then, operating on Eq.\ (\ref{2.4}) by
$ \mathop{\lim}_{\varepsilon \to 0}
      \int_{-\varepsilon}^{\varepsilon} ( \cdot ) \, ds 
$ 
and using $\psi = 0$ for $s < 0$, we obtain
   \begin{align}
      \psi (\bm x, \bm \Omega, 0) = & \; c \int_{4\pi} \int_0^{\infty} 
         P(\bm \Omega' \cdot \bm \Omega) \Sigma_t(\bm\Omega', s') \psi(\bm x, \bm
         \Omega',s') \, ds' d \Omega'  + \frac{Q(\bm x)}{4\pi} \,.
   \end{align}
   \end{subequations}
 Equations (\ref{2.5}) are mathematically equivalent to Eq.\ (\ref{2.4}). 

 To establish the relationship between the present work and the classic number density and angular flux, we integrate Eq.\ (\ref{2.1a}) over $s$ and obtain:
   \begin{align}
   \begin{array}{l}
   \displaystyle{
      \left[ \int_0^{\infty}  n(\bm x, \bm \Omega, s) \, ds \right]
         dV d\Omega} = \\ \\ \end{array} & \begin{array}{l}
         \text{the total number of particles} \\%\vspace{-0.4cm} \\
       \text{in $dV d\Omega$ about $(\bm x, \bm \Omega)$} \;\;.
       \end{array}
   \end{align}
Therefore,
   \begin{align}
      n_c(\bm x, \bm \Omega) = \int_0^{\infty}  n(\bm x, \bm \Omega, s) 
         \, ds = \text{ classic number density, } 
   \end{align}
and
   \begin{align}
      \psi_c(\bm x, \bm \Omega) =
 v n_c(\bm x, \bm \Omega) =  \int_0^{\infty} \psi(\bm x, \bm \Omega,
         s) \, ds = \text{ classic angular flux. }
   \label{2.8}
   \end{align}
   
%%%%%%%%%%%%%%%%%%%%%%%%%%%%%%%%%%%%%%%%%%%%%%%%%

\vspace{20pt}
\noindent
{\bf 3. THE ANGULAR-DEPENDENT PATH-LENGTH AND EQUILIBRIUM PATH-LENGTH DISTRIBUTIONS}
\setcounter{section}{3}
\setcounter{equation}{0} 
\vspace{10pt}

Without loss of generality, let us consider a single particle released from an interaction
site at $x=0$ in the direction ${\bm \Omega} = {\bm i}$ = direction of the
positive $x$-axis. Eq.\ (\ref{2.5a}) for this particle becomes:
  \begin{equation}
    \frac{\partial\psi}{\partial s} (x,\bm\Omega=\bm i, s) + \frac{\partial\psi}{\partial x}
    (x,\bm\Omega=\bm i,s) + \Sigma_t(\bm\Omega=\bm i,s) \psi(x,\bm\Omega=\bm i,s) = 0 \,.
  \label{3.1}
  \end{equation}
For this particle, we have
    $x(s)=s$
    and  $\psi(x(s),\bm i,s) \equiv F(\bm i,s)$. %\nonumber
  Therefore,
  \begin{align}
    \frac{dF}{ds}(\bm i,s) = & \frac{\partial \psi}{\partial x}(x(s),\bm i,s)
      \left( \frac{dx}{ds} \right) +
      \frac{\partial \psi}{\partial s}(x(s),\bm i,s) \\
    = & \frac{\partial \psi}{\partial x} + \frac{\partial \psi}{\partial s}
    \;\;. \nonumber
  \end{align}
Equation (\ref{3.1}) then simplifies to:
  \begin{subequations} \label{3.3}
  \begin{equation}
    \frac{dF}{ds}(\bm i,s) + \Sigma_t(\bm i,s) F(\bm i,s) = 0 \;. %\label{4.3a}
  \end{equation}
We apply the initial condition
  \begin{equation}
    F(\bm i,0)=1 \;,
  \end{equation}
  \end{subequations}
since we are considering a single particle. The solution of Eqs.\ (\ref{3.3})
is:
  \begin{align}
   \begin{array}{l}
    F(\bm\Omega=\bm i,s) = \end{array} & \; e^{-\int_0^s \Sigma_t(\bm\Omega=\bm i,s') \; ds'} \\
    \begin{array}{l} = \\ \\ \end{array}  & \begin{array}{l}
     \text{the probability that the particle will travel the distance}\\%\vspace{-0.4cm} \\
     \text{$s$ in the given direction $\bm\Omega=\bm i$ without interacting.}
  \end{array}
  \nonumber 
  \end{align}
We can generalize this equation for all directions, giving
  \begin{align}
   \begin{array}{l} F(\bm\Omega,s) = \end{array} & \; e^{-\int_0^s \Sigma_t(\bm\Omega,s') \; ds'} \\
    \begin{array}{l} = \\ \\ \end{array} & \begin{array}{l}\text{the probability that the particle will travel the distance} \\%\vspace{-0.4cm} \\
      \text{$s$ in a given direction $\bm\Omega$ without interacting.}
      \end{array}
      \nonumber  
  \end{align}
  
The probability of a collision between $s$ and $s+ds$ in a given direction $\bm\Omega$ is:
  \begin{equation}
    \Sigma_t(\bm\Omega,s) F(\bm\Omega,s) ds = q(\bm\Omega,s) ds \;,
  \end{equation}
and therefore:
  \begin{align}
    \begin{array}{l}
    q(\bm\Omega,s) = \end{array}& \; \Sigma_t(\bm\Omega,s) e^{-\int_0^s \Sigma_t(\bm\Omega,s') \; ds'}\label{3.7}\\
     \begin{array}{l} = \\ \\ \end{array} & \begin{array}{l}
     \text{conditional distribution function for the distance-to-collision}\\%\vspace{-0.4cm}\\
      \text{in a given direction $\bm\Omega$.}
      \end{array}\nonumber
  \end{align}
  
  Let us define 
\begin{equation}
\xi(\bm\Omega)d\Omega = \text{probability that a particle is traveling in $d\Omega$ about $\bm\Omega$;}
\end{equation}
and
\begin{equation}
\begin{array}{l} p(\bm\Omega,s)d\Omega ds = \\ \\\end{array} \begin{array}{l}
\text{probability that a particle traveling in $d\Omega$ about $\bm\Omega$ will}\\%\vspace{-0.4cm}
\text{experience a collision between $s$ and $s+ds$.}
\end{array}
\end{equation}
Then
\begin{align}
p(\bm\Omega,s)d\Omega ds &= \begin{array}{l}
 \text{(prob. that a particle is traveling in $d\Omega$ about $\bm\Omega$)$\times$(prob. of a}\\%\vspace{-0.4cm}\\
\text{collision between $s$ and $s+ds$ in a given direction $\bm\Omega$)}\vspace{-0.4cm}
\end{array}\\
& = (\xi(\bm\Omega)d\Omega )(q(\bm\Omega,s)ds) ;\nonumber
\end{align}
that is, $p(\bm\Omega,s)$ is a joint distribution function.% \cite{grimmett_01}.

 Equation (\ref{3.7}) expresses $q(\bm\Omega,s)$ in terms of $\Sigma_t(\bm\Omega,s)$. To express $\Sigma_t(\bm\Omega,s)$ in terms of $q(\bm\Omega,s)$, we operate on Eq.\ (\ref{3.7}) by $\int_0^s ( \cdot ) ds'$ and get:
  \begin{subequations}
  \begin{align}
 \int_0^s q(\bm\Omega,s') ds' = 1 - e^{-\int_0^s \Sigma_t(\bm\Omega,s') \; ds'} \;\;,
 \end{align}
or
\begin{align}
   e^{-\int_0^s \Sigma_t(\bm\Omega,s') \; ds'} = 1 - \int_0^s q(\bm\Omega,s') ds' \;\;.
   \end{align}
Hence,
  \begin{align} \int_0^s \Sigma_t(\bm\Omega,s') \; ds' = - \ln \left( 1 - \int_0^s q(\bm\Omega,s') ds' \right)
    \;\;. 
    \end{align}
    \end{subequations}
Differentiating with respect to $s$, we obtain:
   \begin{equation} 
      \Sigma_t(\bm\Omega,s) = \frac{q(\bm\Omega,s)}{1 - \int_0^s q(\bm\Omega,s') ds'} \,. 
    \label{3.12}
    \end{equation}
Equations (\ref{3.7}) and (\ref{3.12}) show that $q(\bm\Omega,s)$ is exponential if and only if $\Sigma_t(\bm\Omega,s)$ is
independent of $s$. 

 Moreover, for the case of an infinite medium with an ``equilibrium" intensity having no space dependence (but
dependent on direction), Eq.\ (\ref{2.5a}) for $s > 0$ reduces to:
   \begin{equation}
      \frac{\partial\psi}{\partial s}(\bm\Omega,s) + \Sigma_t(\bm\Omega,s) \psi(\bm\Omega,s) = 0\,,
   \end{equation}
which has the solution
   \begin{equation}
      \psi(\bm\Omega,s) = \psi(\bm\Omega,0) e^{-\int_0^s \Sigma_t(\bm\Omega,s') \; ds'} \,.
   \end{equation}
Normalizing this solution to have integral = unity, we obtain:
   \begin{align}
      \chi(\bm\Omega,s) & = \frac{e^{-\int_0^s \Sigma_t(\bm\Omega,s') \; ds'}}
      {\int_0^{\infty} e^{-\int_0^{s'} \Sigma_t(\bm\Omega,s'') ds''} ds' } \label{3.15}\\
      & = \text{ ``equilibrium" spectrum of path-length $s$ in a given direction $\bm\Omega$ . }\nonumber 
      \end{align}

 From Eq.\ (\ref{3.7}), the mean distance-to-collision (mean free path) in a given direction $\bm\Omega$ is:
   \begin{align}
     s_{\bm\Omega}(\bm\Omega) & = \int_0^{\infty} s q(\bm\Omega,s) \, ds \label{3.16} \\
     & = \int_0^{\infty} s \left[ \Sigma_t(\bm\Omega,s) e^{- \int_0^s \Sigma_t(\bm\Omega,s') ds'} \right] \, ds \nonumber \\
     & = s \left[ - e^{- \int_0^s \Sigma_t(\bm\Omega,s') ds'} \right]_0^{\infty}
          - \int_0^{\infty} \left[ - e^{- \int_0^s \Sigma_t(\bm\Omega,s') ds'} \right] \, ds \nonumber \\
     & = \int_0^{\infty} e^{- \int_0^s \Sigma_t(\bm\Omega,s') ds'} \, ds \,.
     \nonumber
   \end{align}
 Equation (\ref{3.15}) can then be written:
   \begin{equation}
      \chi(\bm\Omega,s) = \frac{e^{- \int_0^s \Sigma_t(\bm\Omega,s') ds'}}{s_{\bm\Omega}(\bm\Omega)} \,,
   \label{3.17}   
   \end{equation}
and by the Law of Total Expectation \cite{billingsley_95}, the mean free path $\bl s \bg$ is given by
\begin{equation}\label{3.18}
\bl s \bg = \int_{4\pi}\int_0^{\infty}sp(\bm\Omega,s) dsd\Omega
=\int_{4\pi}\xi(\bm\Omega) s_{\bm\Omega}(\bm\Omega)d\Omega \; .
\end{equation}
    
 \noindent\textit{Note}: from now on we assume that $s_{\bm\Omega}(\bm\Omega)$ is an \textit{even} function of the direction of flight $\bm\Omega$. This makes sense since, from the physical point of view, the mean free path of a particle traveling in the direction $\bm\Omega$ \textit{must} be equal to the mean free
path of a particle traveling in the direction $-\bm\Omega$.

%%%%%%%%%%%%%%%%%%%%%%%%%%%%%%%%%%%%%%%%%%%%%%%%%

\vspace{20pt}
\noindent
{\bf 4. INTEGRAL EQUATION FORMULATIONS OF THE NEW GLBE}
\setcounter{section}{4}
\setcounter{equation}{0} 
\vspace{10pt}

   Using the work in \cite{larsen_11} as a guide, let us define:
   \begin{align}
       f(\bm x, \bm \Omega) & =  \int_0^{\infty} \Sigma_t(\bm\Omega,s) \psi
         (\bm x, \bm \Omega, s) \, ds \label{4.1} %\\&
      = \text{ collision rate density, } 
   \end{align}
and
   \begin{align}
       g(\bm x, \bm \Omega) & =  \; c \int_{4 \pi} P(\bm \Omega' \cdot \bm
         \Omega)  f(\bm x, \bm \Omega') \, d \Omega'  \label{4.2} %\\&
      =  \text{ inscattering rate density. }
   \end{align}

   The definition (\ref{4.1}) allows us to rewrite Eqs.\ (\ref{2.5}) as:
   \begin{subequations} \label{4.3}
   \begin{gather}
      \frac{\partial \psi}{\partial s}(\bm x, \bm \Omega, s)
         + \bm \Omega \cdot \bm \nabla \psi (\bm x, \bm \Omega, s)
         + \Sigma_t(\bm\Omega,s) \psi (\bm x, \bm \Omega, s) = 0 \,,
         \label{4.3a} \\
      \psi (\bm x, \bm \Omega, 0) = \; c \int_{4\pi}  
         P(\bm \Omega' \cdot \bm \Omega) f(\bm x, \bm
         \Omega') \, d \Omega' + \frac{Q(\bm x)}{4\pi} \,.
   \label{4.3b}
   \end{gather}
   \end{subequations}
Solving Eq.\ (\ref{4.3a}) and using Eq.\ (\ref{4.3b}), we obtain for $s>0$
   \begin{align}
      \psi(\bm x, \bm \Omega, s) &=  \; \psi (\bm x - s \bm \Omega, \bm
         \Omega, 0) e^{- \int_0^s \Sigma_t(\bm\Omega,s') ds'}\label{4.4}  \\
      &\hspace{-0.4cm}= \left[ c \int_{4\pi}  
         P(\bm \Omega' \cdot \bm \Omega)f(\bm x - s \bm \Omega, \bm
         \Omega') \, d \Omega' + \frac{Q(\bm x - s \bm \Omega)}{4\pi}
         \right] e^{- \int_0^s \Sigma_t(\bm\Omega,s') ds'} \,.
   \nonumber
   \end{align}
Operating on this equation by $\int_0^{\infty} \Sigma_t(\bm\Omega,s) (\cdot )
ds$ and using Eqs.\ (\ref{4.1}) and (\ref{3.7}), we get:
   \begin{subequations} \label{4.5}
   \begin{align}
      & f(\bm x, \bm \Omega)\label{4.5a}
      \\& \quad\quad = \int_0^{\infty} \left[ c \int_{4\pi}  
         P(\bm \Omega' \cdot \bm \Omega)  f(\bm x - s \bm \Omega, \bm
         \Omega') \, d \Omega' + \frac{Q(\bm x - s \bm \Omega)}{4\pi}
         \right] q(\bm\Omega,s) \, ds \,.\nonumber
      \end{align}
Also, operating on Eq.\ (\ref{4.4}) by $\int_0^{\infty}  (\cdot ) ds$ and using Eq.\ (\ref{2.8}), we obtain:
   \begin{align}
      &\psi_c(\bm x, \bm \Omega)\label{4.5b}
      \\& = \int_0^{\infty} \left[ c \int_{4\pi}  
         P(\bm \Omega' \cdot \bm \Omega)  f(\bm x - s \bm \Omega, \bm
         \Omega') \, d \Omega' + \frac{Q(\bm x - s \bm \Omega)}{4\pi}
         \right] e^{- \int_0^s \Sigma_t(\bm\Omega,s') ds'} \, ds \,.
   \nonumber
   \end{align}
   \end{subequations}
Thus, if the integral equation (\ref{4.5a}) for $f(\bm x,\bm\Omega)$ is solved, Eq.\ (\ref{4.5b}) yields the classic angular flux.

   Using definition (\ref{4.2}), we can rewrite Eq.\ (\ref{4.5a}) as:
   \begin{equation}
       f(\bm x, \bm \Omega) = \int_0^{\infty} \left[ 
          g(\bm x - s \bm \Omega, \bm \Omega) + \frac{Q(\bm x - s \bm
         \Omega)}{4 \pi} \right] q(\bm\Omega,s) ds \,,
   \label{4.6}
   \end{equation}
and operating on this result by $ c \int_{4\pi} P(\bm \Omega \cdot \bm \Omega' ) ( \cdot ) d \Omega' $ we obtain:
   \begin{equation}
      g(\bm x, \bm \Omega) = c \int_{4\pi} P(\bm \Omega \cdot \bm
         \Omega' ) \int_0^{\infty} \left[ 
          g(\bm x - s \bm \Omega', \bm \Omega') + \frac{Q(\bm x - s \bm
         \Omega')}{4 \pi} \right] q(\bm\Omega',s) \, ds d \Omega' \,.
   \label{4.7}
   \end{equation}
Changing the spatial variables from the 3-D spherical $(\bm \Omega',s)$ to the 3-D Cartesian $\bm x'$ defined by
    \begin{equation}
       \bm x' = \bm x - s \bm \Omega' \,,
    \label{4.8}
    \end{equation}
we obtain
       \begin{subequations} \label{4.9}
       \begin{gather}
          s =  | \bm x - \bm x' | \,,
          \label{4.9a} \\
          \bm \Omega' = \frac{ \bm x - \bm x' }{ | \bm x - \bm x' | } \,,
          \label{4.9b} \\
          s^2 ds d\Omega = dV' \,.
          \label{4.9c}
       \end{gather}
       \end{subequations}

Now, we can rewrite Eq.\ (\ref{4.7}) as:
   \begin{subequations} \label{4.10}
    \begin{align}
       & g(\bm x, \bm \Omega)\label{4.10a}
       \\& \hspace{0.1cm}= c \iiint 
       P \left(\frac{ \bm x - \bm x' }{ | \bm x - \bm x' | } \cdot  \bm \Omega\right)
       \left[
        g \left( \bm x'  , \frac{ \bm x - \bm x' }{ | \bm x - \bm x' | } \right)
       + \frac{Q(\bm x')}{4\pi} \right]
       \frac{ \hat q(|\bm x - \bm x'|)}{| \bm x - \bm x' |^2} \, dV' \,,
    \nonumber
    \end{align}
  where $ \hat q(|\bm x - \bm x'|)dV'$ is the conditional probability that, given the direction defined by $\bm x-\bm x'$, a particle moving from a point $\bm x$ to a point lying in $dV'$ about $\bm x'$ will experience a collision.   
Definition (\ref{4.2}) allows us to rewrite Eq.\ (\ref{4.5b}) as well:
    \begin{equation}
       \psi_c(\bm x, \bm \Omega) = \int_0^{\infty} \left[
        g(\bm x - s \bm \Omega, \bm \Omega) +
       \frac{Q\bigl( \bm x - s \bm \Omega)}{4\pi} \right]
       e^{- \int_0^s \Sigma_t(\bm\Omega,s') ds'} \, ds \,.
    \label{4.10b}
    \end{equation}
   \end{subequations}
Here, solving the integral equation (\ref{4.10a}) for $g(\bm x,\bm\Omega)$, Eq.\ (\ref{4.10b}) yields the classic angular flux. This specific formulation does not contain the path-length variable $s$ as an independent variable.

Finally, for the case of isotropic scattering (in which $ P(\bm \Omega' \cdot \bm \Omega) = 1/4\pi$), $ g(\bm x, \bm \Omega)$ in Eq.\ (\ref{4.2})
becomes isotropic:
   \begin{subequations} \label{4.11}
   \begin{equation}
      g(\bm x) = \frac{c}{4\pi} \int_{4\pi}  f(\bm x, \bm \Omega') d
         \Omega' \equiv \frac{c}{4\pi} \hat F(\bm x) \,,
   \label{4.11a}
   \end{equation}
where
   \begin{equation}
      \hat F(\bm x) = \int_{4\pi} f(\bm x, \bm \Omega') d \Omega' = \text{
         scalar collision rate density. }
   \label{4.11b}
   \end{equation}
   \end{subequations}
 Equation\ (\ref{4.10a}) is then reduced to
   \begin{subequations} \label{4.12}
    \begin{equation}
     \hat F(\bm x) = \iiint \left[ c \hat F(\bm x') + Q(\bm x') \right]  
      \frac{ \hat q(|\bm x - \bm x' |)}{4 \pi | \bm x - \bm x' |^2} \, dV' \,;
    \label{4.12a}
    \end{equation}
and using Eq.\ (\ref{4.11a}), we write Eq.\ (\ref{4.10b}) as:
   \begin{equation}
      \psi_c(\bm x, \bm \Omega) = \frac{1}{4\pi} \int_0^{\infty} \bigl[ c \hat F(\bm x - s \bm \Omega) + Q ( \bm x - s
      \bm \Omega ) \bigr] e^{- \int_0^s \Sigma_t(\bm\Omega,s') ds'} ds \,.
   \label{4.12b}
   \end{equation}
Operating on this equation by $\int_{4\pi} ( \cdot ) d \Omega$ and using Eqs.\ (\ref{4.8}) and (\ref{4.9}), we obtain:
   \begin{equation}
      \phi_c(\bm x) = \iiint \bigl[ c \hat F(\bm x') + Q(\bm x')  \bigr] \frac{e^{- \int_0^{| \bm x - \bm x' |} 
         \Sigma_t\left(\frac{\bm x-\bm x'}{|\bm x-\bm x'|},s'\right) ds'} } {4 \pi | \bm x - \bm x' |^2} dV'  \,.
   \label{4.12c}
   \end{equation}   
   \end{subequations}
If we solve the integral equation (\ref{4.12a}) for $\hat F(\bm x)$, the classic angular flux is given by Eq.\ (\ref{4.12b}) and the classic scalar flux is given by Eq.\ (\ref{4.12c}).
  
%%%%%%%%%%%%%%%%%%%%%%%%%%%%%%%%%%%%%%%%%%%%%%%%%

\vspace{20pt}
\noindent
{\bf 5. ASYMPTOTIC DIFFUSION LIMIT OF THE NEW GLBE}
\setcounter{section}{5}
\setcounter{equation}{0} 
\vspace{10pt}

   To begin this discussion, we must first consider the Legendre polynomial expansion of the distribution function
 $P(\bm \Omega \cdot \bm \Omega') = P(\mu_0)$ defined by Eq.\ (\ref{2.1f}) \cite{lewis_93}:
   \begin{equation}
    P(\mu_0) = \sum_{n=0}^{\infty} \frac{2n+1}{4\pi} a_n P_n(\mu_0) \,,
     \end{equation}
where $a_0 = 1$ and $a_1 = \overline \mu_0$ = mean scattering cosine. We define $P^*(\mu_0)$ by:
   \begin{equation}
      P^*(\mu_0) = cP(\mu_0) + \frac{1-c}{4\pi} \,,
   \label{5.2}
   \end{equation}
which has the Legendre polynomial expansion:
   \begin{subequations} \label{5.3}
   \begin{gather}
      P^*(\mu_0) = \sum_{n=0}^{\infty} \frac{2n+1}{4\pi} a_n^* P_n(\mu_0) \,,  \label{5.3a} \\
      a_n^* = \bigg\{\begin{array}{lcl}
      1 & , & n=0 \\%\vspace{-0.3cm}\\
      ca_n & , & n \ge 1 
      \end{array} \,.  \label{5.3b}
   \end{gather}
   \end{subequations}
   
 Using the work in \cite{larsen_96} as a guide, we scale $\Sigma_t = O(\varepsilon^{-1})$, 
$ 1-c = O(\varepsilon^2) $, $Q=O(\varepsilon)$, $P^*(\mu_0)$ is independent of $\varepsilon$, and
 $\partial \psi / \partial s = O(\varepsilon^{-1})$, with $\varepsilon \ll 1$. Equations (\ref{2.4}) and (\ref{5.2}) yield, in this scaling,
  \begin{align}
    \frac{1}{\varepsilon} \frac{\partial \psi}{\partial s} & (\bm x, \bm \Omega,s) 
      + \bm \Omega \cdot \bm \nabla \psi(\bm x, \bm \Omega, s)
       + \frac{\Sigma_t(\bm \Omega,s)}{\varepsilon} \psi(\bm x, \bm \Omega, s)   \label{5.4}\\
   & = \delta(s) \int_{4\pi} \int_0^{\infty} \left[ P^*(\bm \Omega \cdot \bm \Omega' ) 
      - \varepsilon^2 \frac{1-c}{4\pi }\right] \frac{\Sigma_t(\bm \Omega',s')}{\varepsilon} 
      \psi(\bm x, \bm \Omega', s') \, ds' d\Omega' \nonumber\\
      &\quad\quad\quad\quad + \varepsilon \delta(s) \frac{Q(\bm x)}{4\pi} \,.
 \nonumber
  \end{align}

 Let us define $\Psi(\bm x, \bm \Omega, s)$ by:
  \begin{equation}
     \psi(\bm x, \bm \Omega, s) \equiv 
          \Psi(\bm x, \bm \Omega, s) \frac{e^{-\int_0^s \Sigma_t(\bm \Omega,s') ds'}}{\bl s\bg}\,.
  \end{equation}
Then, using Eq.\ (\ref{3.7}), Eq.\ (\ref{5.4}) for $\psi(\bm x, \bm \Omega, s)$ becomes the following equation for $\Psi(\bm x, \bm \Omega, s)$:
  \begin{align}\label{5.6}
    \frac{\partial \Psi}{\partial s} & (\bm x, \bm \Omega,s) 
      + \varepsilon \bm \Omega \cdot \bm \nabla \Psi(\bm x, \bm \Omega, s)  \\
   & = \delta(s) \int_{4\pi} \int_0^{\infty} \left[ P^*(\bm \Omega \cdot \bm \Omega' ) 
      - \varepsilon^2 \frac{1-c}{4\pi } \right] q(\bm\Omega',s')
      \Psi(\bm x, \bm \Omega', s') \, ds' d\Omega'\nonumber\\
      & \quad\quad\quad\quad
      + \varepsilon^2 \delta(s) \bl s\bg \frac{Q(\bm x)}{4\pi} \,.\nonumber 
  \end{align}
The scaling in this equation implies the following:
\begin{itemize}
\item The $O(1)$ terms describe particle scattering. The transport process is dominated by scattering; the length scale for the problem is chosen so that a unit of length is comparable to a typical mean free path, and the system is many mean free paths thick.\vspace{-3pt}
\item The leakage $(\bm\Omega\cdot\nabla\Psi)$ term is $O(\varepsilon)$. (The angular flux $\psi$ varies a small amount over the distance of one mean free path.)\vspace{-3pt}
\item The absorption term $1-c = \Sigma_a/\Sigma_t$ and the source term $Q$ are $O(\varepsilon^2)$, balanced in such a way that the infinite medium solution $\psi=Q/4\pi\Sigma_a$ is $O(1)$. (The infinite medium solution holds when the source and the cross sections are constant.)\vspace{-3pt}
\item In the scaling defined by Eqs.\ (\ref{5.3}) only the $n = 0$ constant $a_0$ is ``stretched" asymptotically; the higher-order	$(n\ge 1)$ terms are not stretched. When this scaling is applied to a standard linear Boltzmann equation one obtains the same diffusion equation that is obtained from the standard $P_1$ or spherical harmonics approximation.
 \end{itemize}
 Equation\ (\ref{5.6}) is mathematically equivalent to:
   \begin{subequations}
   \begin{align}
      &\frac{\partial \Psi}{\partial s} (\bm x, \bm \Omega,s) 
         + \varepsilon \bm \Omega \cdot \bm \nabla \Psi(\bm x, \bm \Omega, s) = 0 \quad , \quad s > 0 \,,\label{5.7a}\\
       &\Psi(\bm x, \bm \Omega, 0)  =\int_{4\pi} \left[ P^*(\bm \Omega \cdot \bm \Omega' ) 
         - \varepsilon^2 \frac{1-c}{4\pi} \right] \int_0^{\infty} q(\bm\Omega',s')
\Psi(\bm x, \bm \Omega', s') ds' d\Omega'\\
         &\hspace{3 cm} + \varepsilon^2 \bl s\bg \frac{Q(\bm x)}{4\pi} \,.\nonumber
   \end{align}
   \end{subequations}
Integrating Eq.\ (\ref{5.7a}) over $0 < s' < s$, we obtain:
   \begin{align}
      \Psi(\bm x, \bm\Omega, s) & = \Psi(\bm x, \bm\Omega, 0) - \varepsilon \bm \Omega \cdot \bm \nabla
         \int_0^s \Psi(\bm x, \bm\Omega, s') \, ds' \\
      & = \int_{4\pi} \left[ P^*(\bm \Omega \cdot \bm \Omega' ) 
         - \varepsilon^2 \frac{1-c}{4\pi} \right] \int_0^{\infty} q(\bm\Omega',s')
\Psi(\bm x, \bm \Omega', s') ds' d\Omega'
      \nonumber \\
      & \quad\quad\quad + \varepsilon^2 \bl s\bg \frac{Q(\bm x)}{4\pi}
         - \varepsilon \bm \Omega \cdot \bm \nabla
         \int_0^s \Psi(\bm x, \bm\Omega, s') \, ds' \,. \nonumber 
   \end{align}
    Introducing into this equation the ansatz 
   \begin{equation}
 \Psi(\bm x, \bm \Omega, s) =  \sum_{n=0}^{\infty} \varepsilon^n
       \Psi^{(n)}(\bm x, \bm \Omega, s) 
       \end{equation}
and equating the coefficients of different powers of $\varepsilon$, we obtain for $n \ge 0$:
   \begin{align}
      \Psi^{(n)}(\bm x, \bm \Omega, s) & = \int_{4\pi} P^*(\bm \Omega \cdot \bm \Omega' ) 
         \int_0^{\infty} q(\bm\Omega',s')
\Psi^{(n)}(\bm x, \bm \Omega', s') ds' d\Omega'
       \label{5.10} \\
      & \quad -  \bm \Omega \cdot \bm \nabla
         \int_0^s \Psi^{(n-1)}(\bm x, \bm\Omega, s') \, ds'  \nonumber  \\
      & \quad \quad - \frac{1-c}{4\pi} \int_{4\pi} \int_0^{\infty} q(\bm\Omega',s')
\Psi^{(n-2)}(\bm x, \bm \Omega', s') ds' d\Omega'
          \nonumber \\
      & \quad \quad \quad + \delta_{n,2} \bl s\bg \frac{Q(\bm x)}{4\pi} \,. 
   \nonumber
   \end{align}
We solve these equations recursively, using the Legendre
 polynomial expansion (\ref{5.2}) of $ P^*(\mu_0) $.

 Equation (\ref{5.10}) with $n=0$ is:
   \begin{equation}
 \Psi^{(0)}(\bm x, \bm \Omega, s) = \int_{4\pi} P^*(\bm \Omega \cdot \bm \Omega' ) 
         \int_0^{\infty} q(\bm\Omega',s')
\Psi^{(0)}(\bm x, \bm \Omega', s') ds' d\Omega' \,. 
\end{equation}
The general solution of this equation is:
   \begin{equation}
      \Psi^{(0)}(\bm x, \bm \Omega, s) = \frac{\Phi^{(0)}(\bm x)}{4\pi} \,,
   \label{5.12}
   \end{equation}
where $\Phi^{(0)}(\bm x)$ is, at this point, undetermined.

   Next, Eq.\ (\ref{5.10}) with $n=1$ is:
   \begin{align}
      \Psi^{(1)}(\bm x, \bm \Omega, s) & = \int_{4\pi} P^*(\bm \Omega \cdot \bm \Omega' ) 
         \int_0^{\infty} q(\bm\Omega',s')
\Psi^{(1)}(\bm x, \bm \Omega', s') ds' d\Omega' \label{5.13} \\
       & \quad\quad\quad  -  s \bm \Omega \cdot \bm \nabla \frac{\Phi^{(0)}(\bm x)}{4\pi} \;\;. 
   \nonumber
   \end{align}
This equation has a particular solution of the form:
    \begin{equation}
      \Psi_{part}^{(1)}(\bm x, \bm \Omega, s) = [\bm\tau(\bm\Omega) - s\bm \Omega]  \cdot \bm \nabla \frac{\Phi^{(0)}(\bm x)}{4\pi} \,,
   \label{5.14}
   \end{equation}   
where 
\begin{subequations}\label{5.15}
\begin{align}
& \bm\tau(\bm\Omega) = \int_{4\pi} P^*(\bm\Omega\cdot\bm\Omega')\bm\tau(\bm\Omega')d\Omega' + \bm{\hat S}(\bm\Omega)\,,\label{5.15a}\\
& \bm{\hat S}(\bm\Omega) = -\int_{4\pi} \bm\Omega'P^*(\bm\Omega\cdot\bm\Omega')s_{\bm\Omega}(\bm\Omega')d\Omega'\,.
\end{align}
\end{subequations}
As a Fredholm integral equation of the second kind, Eq.\ (\ref{5.15a}) has the Liouville-Neumann series solution \cite{arfken_85}:
\begin{align}\label{5.16}
\bm\tau(\bm\Omega) = \lim_{N\to\infty}\sum_{n=0}^N \bm\tau_n(\bm\Omega)\,,
\end{align}
where
\begin{subequations}\label{5.17}
\begin{align}
\bm\tau_0(\bm\Omega) &= \bm{\hat S}(\bm\Omega)\,,
\\ \bm\tau_1(\bm\Omega) &= \int_{4\pi}P^*(\bm\Omega\cdot\bm\Omega_1)\bm{\hat S}(\bm\Omega_1)d\Omega_1\,,
\\ \bm\tau_2(\bm\Omega) &= \int_{4\pi}\int_{4\pi}P^*(\bm\Omega\cdot\bm\Omega_1)P^*(\bm\Omega_1\cdot\bm\Omega_2)\bm{\hat S}(\bm\Omega_2)d\Omega_2d\Omega_1\,,
\\ & \begin{array}{l}
\cdot \\%\vspace{-.5 cm}\\
 \cdot \\%\vspace{-.5 cm}\\
  \cdot
\end{array}\nonumber
\\ \bm\tau_n(\bm\Omega) &= \int_{4\pi}\int_{4\pi}...\int_{4\pi}P^*(\bm\Omega\cdot\bm\Omega_1)P^*(\bm\Omega_1\cdot\bm\Omega_2)\,...
\\& \quad\quad\quad\quad\quad\quad\quad\quad\quad\quad\quad...\,P^*(\bm\Omega_{n-1}\cdot\bm\Omega_n)\bm{\hat S}(\bm\Omega_n)d\Omega_n\,...\,d\Omega_2d\Omega_1\,.\nonumber
\end{align}
\end{subequations}

\noindent Since $s_{\bm\Omega}(\bm\Omega)$ is an even function of $\bm\Omega$, we note that $\bm{\hat S}(\bm\Omega)$ and $\bm\tau(\bm\Omega)$ are odd functions of $\bm\Omega$:

\begin{align}
\bm{\hat S}(-\bm\Omega) &= -\int_{4\pi} \bm\Omega'P^*(-\bm\Omega \cdot\bm\Omega')s_{\bm\Omega}(\bm\Omega')d\Omega'
\\& = -\int_{4\pi} -\bm\Omega'P^*(-\bm\Omega \cdot-\bm\Omega')s_{\bm\Omega}(-\bm\Omega')d\Omega'\nonumber
\\& = \int_{4\pi} \bm\Omega'P^*(\bm\Omega \cdot\bm\Omega')s_{\bm\Omega}(\bm\Omega')d\Omega' \nonumber
\\& =-\bm{\hat S}(\bm\Omega)\,,\nonumber
\end{align}
and $\bm\tau_n(-\bm\Omega) = -\bm\tau_n(\bm\Omega)$ $\forall$ $n\ge0$ follows with the same argument, by Eqs. (\ref{5.17}).
 The general solution of Eq.\ (\ref{5.13}) is given by:
   \begin{equation}
      \Psi^{(1)}(\bm x, \bm \Omega, s) =  \frac{\Phi^{(1)}(\bm x)}{4\pi} + [\bm\tau(\bm\Omega)- s\bm \Omega]  \cdot \bm \nabla \frac{\Phi^{(0)}(\bm x)}{4\pi} \,,
   \label{5.19}
  \end{equation}  
where $\Phi^{(1)}(\bm x)$ is undetermined.

 Equation (\ref{5.10}) with $n=2$ has a solvability condition, which is obtained by
 operating on it by $ \int_{4\pi} \int_0^{\infty} q(\bm\Omega,s) ( \cdot ) ds d\Omega $.
 Using Eqs.\ (\ref{5.12}) and (\ref{5.19}) to obtain:
   \begin{subequations}
   \begin{align}
 \int _{4\pi}  \int_0^{\infty} q(\bm\Omega',s')\Psi^{(0)}(\bm x, \bm \Omega', s')\, ds' d\Omega'
 = \Phi^{(0)}(\bm x)    \,, 
      \end{align}
and:   
   \begin{align}
 \int_0^s \Psi^{(1)}(\bm x, \bm \Omega, s') \, ds' = s\frac{\Phi^{(1)}(\bm x)}{4\pi}
       + \left(s\bm\tau(\bm\Omega)- \frac{s^2}{2}\bm \Omega\right)   \cdot \bm \nabla \frac{\Phi^{(0)}(\bm x)}{4\pi} \,, 
       \end{align}
      \end{subequations}
the solvability condition becomes:
   \begin{align}
      &0 = \frac{1}{4\pi} \int_{4\pi}\int_0^{\infty}q(\bm\Omega,s)\left(\frac{s^2}{2}[\bm \Omega \cdot \bm \nabla]^2-s[\bm\tau(\bm\Omega)\cdot\bm\nabla][\bm\Omega\cdot\bm\nabla]\right)  
          \Phi^{(0)}(\bm x) ds d\Omega \label{5.21}\\
          & \quad\quad\quad\quad\quad\quad\quad\quad - \frac{(1-c)}{4\pi} \int_{4\pi} \int_0^{\infty} q(\bm\Omega,s) \Phi^{(0)}(\bm x) \, ds d\Omega 
         + \bl s\bg Q(\bm x)\,.\nonumber 
   \end{align}
Thus, using the fact that $\int_0^{\infty}q(\bm\Omega,s)ds = 1$ and
$\int_0^{\infty}s^mq(\bm\Omega,s)ds = s^m_{\bm\Omega}(\bm\Omega)$, % and dividing it by $\overline s$,
we can rewrite Eq.\ (\ref{5.21}) as:
\begin{align}\label{5.22}
      \frac{1}{4\pi\bl s\bg} \int_{4\pi}\left( \frac{s_{\bm\Omega}^2(\bm\Omega)}{2}[\bm \Omega \cdot  \bm\nabla]^2-s_{\bm\Omega}(\bm\Omega)[\bm\tau(\bm\Omega)\cdot\bm\nabla][\bm\Omega\cdot\bm\nabla]\right)& \Phi^{(0)}(\bm x) \, d\Omega %\nonumber \\
  \\&   \hspace{-1.5 cm}- \frac{(1-c)}{\bl s\bg } \Phi^{(0)}(\bm x)  
         +  Q(\bm x) = 0 \,. \nonumber 
   \end{align}
If we write $\bm\tau(\bm\Omega) = \bm(\tau_x(\bm\Omega),\tau_y(\bm\Omega),\tau_z(\bm\Omega)\bm)$, this equation is equivalent to:
\begin{align}\label{5.23}
   -\bigg[\text{D}_{xx} &\frac{\partial^2}{\partial x^2} + \text{D}_{yy}\frac{\partial^2}{\partial y^2}
   +\text{D}_{zz}\frac{\partial^2}{\partial z^2}
   \\& + \text{D}_{xy}\frac{\partial^2}{\partial x\partial y} + \text{D}_{xz}\frac{\partial^2}{\partial x\partial z}
   + \text{D}_{yz}\frac{\partial^2}{\partial y\partial z}\bigg]\Phi^{(0)}(\bm x)
+ \frac{1-c}{\bl s\bg} \Phi^{(0)}(\bm x) = Q(\bm x)\,,\nonumber
\end{align}
where D$_{xx}$, D$_{yy}$, D$_{zz}$, D$_{xy}$, D$_{xz}$, and D$_{yz}$ are the diffusion coefficients given by
\begin{subequations}\label{5.24}
\begin{align}
\text{D}_{xx} &= \frac{1}{4\pi\bl s\bg} \int_{4\pi} \left( \frac{s^2_{\bm\Omega}(\bm\Omega)}{2}\Omega_x-s_{\bm\Omega}(\bm\Omega)\tau_x(\bm\Omega)\right)\Omega_xd\Omega,\\
\text{D}_{yy} &= \frac{1}{4\pi\bl s\bg} \int_{4\pi} \left( \frac{s^2_{\bm\Omega}(\bm\Omega)}{2}\Omega_y-s_{\bm\Omega}(\bm\Omega)\tau_y(\bm\Omega)\right)\Omega_yd\Omega,\\
\text{D}_{zz} &= \frac{1}{4\pi\bl s\bg} \int_{4\pi} \left( \frac{s^2_{\bm\Omega}(\bm\Omega)}{2}\Omega_z-s_{\bm\Omega}(\bm\Omega)\tau_z(\bm\Omega)\right)\Omega_zd\Omega,\\
\text{D}_{xy} &= \frac{1}{4\pi\bl s\bg} \int_{4\pi} \left( s^2_{\bm\Omega}(\bm\Omega)\Omega_x\Omega_y-s_{\bm\Omega}(\bm\Omega)[\tau_x(\bm\Omega)\Omega_y+\tau_y(\bm\Omega)\Omega_x]\right)d\Omega,\\
\text{D}_{xz} &= \frac{1}{4\pi\bl s\bg} \int_{4\pi} \left( s^2_{\bm\Omega}(\bm\Omega)\Omega_x\Omega_z-s_{\bm\Omega}(\bm\Omega)[\tau_x(\bm\Omega)\Omega_z+\tau_z(\bm\Omega)\Omega_x]\right)d\Omega,\\
\text{D}_{yz} &= \frac{1}{4\pi\bl s\bg} \int_{4\pi} \left( s^2_{\bm\Omega}(\bm\Omega)\Omega_y\Omega_z-s_{\bm\Omega}(\bm\Omega)[\tau_y(\bm\Omega)\Omega_z+\tau_z(\bm\Omega)\Omega_y]\right)d\Omega.
\end{align}
\end{subequations}

Summarizing: the solution $\psi(\bm x, \bm \Omega, s)$ of Eq.\ (\ref{5.4}) satisfies:
   \begin{equation}
      \psi(\bm x, \bm \Omega, s) = \frac{\Phi^{(0)}(\bm x)}{4\pi} \frac{e^{- \int_0^s \Sigma_t(\bm\Omega, s') ds'}} 
         {\bl s\bg} + O(\varepsilon) \,,
   \label{5.25}
   \end{equation} 
where $\Phi^{(0)}(\bm x)$ satisfies Eq.\ (\ref{5.23}). Integrating Eq.\ (\ref{5.25}) over $0 < s < \infty$
and using equation (\ref{3.16}), we obtain an expression
 to the classic angular flux (to leading order):
   \begin{align}
   \psi_c(\bm x, \bm \Omega) = \Phi^{(0)}(\bm x)\frac{s_{\bm\Omega}(\bm\Omega)}{4\pi\bl s\bg} \,.
   \label{5.26}
   \end{align}

\vspace{20pt}
\noindent
{\bf 5.1 Special Case 1: Isotropic Scattering}
\vspace{10pt}

In the case of isotropic scattering, $P^*(\bm\Omega \cdot \bm\Omega') = 1/4\pi$ and
\begin{align}
 \bm{\hat S}(\bm\Omega)= -\frac{1}{4\pi}\int_{4\pi} \bm\Omega'\,s_{\bm\Omega}(\bm\Omega')d\Omega' = 0\,,
 \end{align}
since $s_{\bm\Omega}(\bm\Omega)$ is an even function of $\bm\Omega$.
 Introducing this result into Eqs.\ (\ref{5.17}), we obtain $\bm\tau_n(\bm\Omega) = 0$ $\forall$ $n\ge 0$. Hence, by Eq.\ (\ref{5.16}), $\bm\tau(\bm\Omega) = 0$, and Eqs.\ (\ref{5.24}) can be writen as
\begin{subequations}\label{5.28}
\begin{align}
\text{D}_{xx} &= \frac{1}{2\bl s\bg}\left(\frac{1}{4\pi}\int_{4\pi} s^2_{\bm\Omega}(\bm\Omega)\Omega_x^2d\Omega\right),\label{5.28a}\\
\text{D}_{yy} &= \frac{1}{2\bl s\bg}\left(\frac{1}{4\pi} \int_{4\pi} s^2_{\bm\Omega}(\bm\Omega)\Omega_y^2d\Omega\right),\label{5.28b}\\
\text{D}_{zz} &= \frac{1}{2\bl s\bg}\left(\frac{1}{4\pi} \int_{4\pi} s^2_{\bm\Omega}(\bm\Omega)\Omega_z^2d\Omega\right),\label{5.28c}\\
\text{D}_{xy} &= \frac{1}{\bl s\bg}\left(\frac{1}{4\pi}\int_{4\pi} s^2_{\bm\Omega}(\bm\Omega)\Omega_x\Omega_yd\Omega\right),\label{5.28d}\\
\text{D}_{xz} &= \frac{1}{\bl s\bg}\left(\frac{1}{4\pi}\int_{4\pi} s^2_{\bm\Omega}(\bm\Omega)\Omega_x\Omega_z d\Omega\right),\label{5.28e}\\
\text{D}_{yz} &= \frac{1}{\bl s\bg}\left(\frac{1}{4\pi}\int_{4\pi} s^2_{\bm\Omega}(\bm\Omega)\Omega_y\Omega_zd\Omega\right)\label{5.28f}.
\end{align}
\end{subequations}

\vspace{20pt}
\noindent
{\bf 5.2 Special Case 2: Isotropic Scattering and Azimuthal Symmetry}
\vspace{10pt}

A general diffusion equation with no off-diagonal terms (that is, without diffusion coefficients that depend on more than one direction)
can be obtained in systems with azimuthal symmetry (such as in PBR problems). Specifically, if we define the polar angle against the $z$-axis such that $(\Omega_x,\Omega_y,\Omega_z) = (\sqrt{1-\mu^2}\cos\varphi,\sqrt{1-\mu^2}\sin\varphi,\mu)$,
the probability distribution function for distance-to-collision is independent of the azimuthal angle $\varphi$; that is, $s_{\bm\Omega}^m(\bm\Omega) = s_{\bm\Omega}^m(\mu)$ depends only upon the polar angle $\mu$.
 Then, 
  D$_{xy}=\text{D}_{xz}=\text{D}_{yz}=0$ and
$\text{D}_{xx}=\text{D}_{yy}$, and we obtain the following anisotropic diffusion equation for $\Phi^{(0)}(\bm x)$:
   \begin{equation}\label{c36.12}
   -\text{D}_{xx} \frac{\partial^2}{\partial x^2}\Phi^{(0)}(\bm x) - \text{D}_{yy}\frac{\partial^2}{\partial y^2}\Phi^{(0)}(\bm x)
   -\text{D}_{zz}\frac{\partial^2}{\partial z^2}\Phi^{(0)}(\bm x)
+ \frac{1-c}{\bl s\bg} \Phi^{(0)}(\bm x) = Q(\bm x),
\end{equation}
where D$_{xx}=\text{D}_{yy}$ are given by Eq.\ (\ref{5.28a}) $=$ Eq.\ (\ref{5.28b}), and D$_{zz}$ is given by Eq.\ (\ref{5.28c}).

\vspace{20pt}
\noindent
{\bf 5.3 Special Case 3:  Standard GLBE ($\Sigma_t(\bm\Omega,s) = \Sigma_t(s)$)}
\vspace{10pt}
 
 Let us now examine the situation in which the locations of the scattering centers are correlated but independent of direction. In this case, we can write $\Sigma_t(\bm\Omega,s) = \Sigma_t(s)$,
and Eq.\ (\ref{3.7}) yields
$q(\bm\Omega,s) = \Sigma_t(s)e^{-\int_0^s\Sigma_t(s')ds'}$.
Introducing this result into
 Eq.\ (\ref{3.16}), we see that $s_{\bm\Omega}(\bm\Omega) = s_{\bm\Omega}$ is now independent of $\bm\Omega$, and we can use Eq.\ (\ref{3.18}) to obtain $\bl s \bg= s_{\bm\Omega}$. $\big($Similarly,
$s_{\bm\Omega}^2(\bm\Omega)=\bl s^2\bg$.$\big)$ Therefore,
 operating on Eq.\ (\ref{5.26})
by $\int_{4\pi}(\cdot)d\Omega$, we obtain
\begin{align}
 \int_{4\pi}\psi_c(\bm x,\bm\Omega)d\Omega = \Phi^{(0)}(\bm x)\,.
 \end{align}
Thus, the solution $\Phi^{(0)}(\bm x)$ of Eq.\ (\ref{5.23}) is the classic scalar flux (to leading order). Furthermore, we can write   
\begin{subequations}\label{5.31}
\begin{align}
 \bm{\hat S}(\bm\Omega)= -\int_{4\pi} \bm\Omega'P^*(\bm\Omega\cdot\bm\Omega')s_{\bm\Omega}(\bm\Omega')d\Omega' = 
 -\bl s\bg\int_{4\pi} \bm\Omega'P^*(\bm\Omega\cdot\bm\Omega')d\Omega'.
\end{align}
To evaluate this integral, we choose the system of coordinates such that \linebreak
$\bm\Omega  = (0,0,1)=\bm{\vec{k}}$. Then, $\bm\Omega\cdot\bm\Omega' = \mu'$ and
\begin{align}
\int_{4\pi} \bm\Omega'P^*(\bm\Omega\cdot\bm\Omega')d\Omega' = 2\pi\bm{\vec{k}}\int_{-1}^1\mu'P^*(\mu')d\mu'
=2\pi\bm\Omega\int_{-1}^1\mu'P^*(\mu')d\mu'.
\end{align}
We know \cite{lewis_93} that $P_1(\mu')=\mu'$; thus, using Eqs.\ (\ref{5.3}):
\begin{align}
\int_{-1}^1\mu'P^*(\mu')d\mu' = \int_{-1}^1\frac{3}{4\pi}a_1^*\mu'^2d\mu' = \frac{a_1^*}{2\pi}= \frac{ca_1}{2\pi},
\end{align}
\end{subequations}
due to the orthogonality of the Legendre polynomials. Since $a_1 = \overline \mu_0$ (the mean scattering cosine), Eqs.\ (\ref{5.31}) yield the explicit expression
\begin{align}
\bm{\hat S}(\bm\Omega) = -\bl s\bg \left( 2\pi\bm\Omega\frac{c\overline\mu_0}{2\pi}\right) = -\bl s\bg\,\bm\Omega [c\overline\mu_0]\,.
\end{align}
Introducing this equation into Eqs.\ (\ref{5.17}), we obtain $\bm\tau_n(\bm\Omega) =  -\bl s\bg \,\bm\Omega[c\overline\mu_0]^{n+1}$ $\forall$ $n\ge 0$; and using Eq.\ (\ref{5.16}): 
\begin{align}
\bm\tau(\bm\Omega)=
 - \frac{c\overline\mu_0}{1-c\overline\mu_0}\,\bl s\bg\,\bm\Omega\,.
\end{align} 
In this case, the angular integrals in Eqs.\ (\ref{5.24}) yield: 
\begin{subequations}\label{5.34}
\begin{align}
&\text{D}=\text{D}_{xx} = \text{D}_{yy} = \text{D}_{zz}= \frac{1}{3}\left(\frac{\bl s^2\bg}{2\bl s\bg}+\frac{c\overline\mu_0}{1-c\overline\mu_0}\bl s\bg\right)\,,
\\& \text{D}_{xy} = \text{D}_{xz} = \text{D}_{yz}= 0 \,,
\end{align}
\end{subequations}
which reduces Eq.\ (\ref{5.23}) to the result obtained in \cite{larsen_07}. 
\\

\noindent {\em Note}: if $q(\bm\Omega,s)$ were to decay algebraically as $s\longrightarrow\infty$ as:
\begin{align}
q(\bm\Omega,s)\geq \frac{constant}{s^3} \quad \text{for} \quad s\geq 1\,,
\end{align}
then the asymptotic diffusion approximation developed here would be invalid, since this would imply
\begin{align}
s_{\bm\Omega}^2(\bm\Omega) = \int_0^{\infty}s^2q(\bm\Omega,s)ds = \infty\,.
\end{align}
The asymptotic analysis tacitly requires both $s_{\bm\Omega}$ and $s^2_{\bm\Omega}$ to be finite. Physically, when $s^2_{\bm\Omega}=\infty$, particles will travel large distances between collisions too often. That is, sufficiently long flight paths will occur sufficiently often that the diffusion description developed here becomes invalid. Following the nomenclature in \cite{davis_08}, standard diffusion ($s^2_{\bm\Omega}<\infty$) is an asymptotic approximation of the GLBE, while anomalous diffusion ($s^2_{\bm\Omega}=\infty$) is not.

%%%%%%%%%%%%%%%%%%%%%%%%%%%%%%%%%%%%%%%%%%%%%%%%%

\vspace{20pt}
\noindent
{\bf 6. REDUCTION TO THE CLASSIC THEORY}
\setcounter{section}{6}
\setcounter{equation}{0} 
\vspace{10pt}

   We now show that, with the classic assumption that the locations of the scattering centers are uncorrelated and do not depend upon direction, the results obtained by the GLBE presented here reduce to the results of the classic theory. In other words, we now assume that
\begin{align}\label{6.1}
\Sigma_t(\bm\Omega,s) = \Sigma_t \equiv \text{constant.} 
\end{align}
In this case, Eq.\ (\ref{2.4}) can be rewritten as
 \begin{align}
      \frac{\partial \psi}{\partial s}(\bm x, \bm \Omega, s)&
          + \bm \Omega \cdot \bm \nabla \psi (\bm x, \bm \Omega, s)
         + \Sigma_t \psi (\bm x, \bm \Omega, s) \\
      & = \delta(s) \, \Sigma_s \int_{4\pi} \int_0^{\infty} P(\bm \Omega' \cdot
         \bm \Omega) \psi(\bm x, \bm \Omega',s') \, ds' 
         d \Omega' + \delta(s) \frac{Q(\bm x)}{4\pi} \,, \nonumber
   \end{align}
where $\Sigma_s = c\Sigma_t$. Operating on this equation by $\int_{-\varepsilon}^{\infty}(\cdot)ds$ and using Eq.\ (\ref{2.8}),
we obtain 
\begin{align}
      \psi(\bm x, \bm \Omega, \infty)-\psi(\bm x, \bm \Omega, -\varepsilon)
          + & \bm \Omega \cdot \bm \nabla \psi_c (\bm x, \bm \Omega)
         + \Sigma_t \psi_c (\bm x, \bm \Omega)\\ & = \Sigma_s \int_{4\pi} P(\bm \Omega' \cdot
         \bm \Omega) \psi_c(\bm x, \bm \Omega') \, 
         d \Omega' +  \frac{Q(\bm x)}{4\pi} \,. \nonumber
   \end{align}
Using the fact that $\psi(\bm x, \bm \Omega, \infty)=\psi(\bm x, \bm \Omega, -\varepsilon)=0$, we have
\begin{align}
\bm \Omega \cdot \bm \nabla \psi_c (\bm x, \bm \Omega)
         + \Sigma_t \psi_c (\bm x, \bm \Omega) = \Sigma_s \int_{4\pi} P(\bm \Omega' \cdot
         \bm \Omega) \psi_c(\bm x, \bm \Omega') \, 
         d \Omega' +  \frac{Q(\bm x)}{4\pi} \,, 
   \end{align}
which is, of course, the classic linear Boltzmann equation. 

Moreover, if Eq.\ (\ref{6.1}) holds, then Eq.\ (\ref{3.7}) yields
\begin{align}\label{6.5}
q(\bm\Omega,s) = \Sigma_te^{-\Sigma_ts} = p(s)\,;
\end{align}
that is, the probability distribution function for distance-to-collision is given by an exponential.
 Introducing Eq.\ (\ref{6.5}) into Eq.\ (\ref{3.16}), we can use Eq.\ (\ref{3.18}) to obtain
\begin{subequations}\label{6.6}
\begin{align}
\bl s\bg = s_{\bm\Omega}(\bm\Omega) = \frac{1}{\Sigma_t}\,, \label{6.6a}
\end{align}
which is the classic expression for the mean free path. Also, the mean-squared free path is given by:
\begin{align}
\bl s^2\bg = s^2_{\bm\Omega}(\bm\Omega) = \int_0^\infty s^2p(s)ds = \frac{2}{\Sigma_t^2}\,.
\end{align}    
\end{subequations} 

 For the integral formulation, Eqs.\ (\ref{4.10}) can now be easily reduced to their classic form,
 since Eq.\ (\ref{6.1}) allows us to write
 \begin{align}\label{6.7}
  \hat q(|\bm x - \bm x'|) = \Sigma_te^{-\Sigma_t|\bm x - \bm x'|}
 \end{align}
 in Eq.\ (\ref{4.10a}). Furthermore, Eq.\ (\ref{4.1}) yields
 \begin{align}
   f(\bm x, \bm \Omega) = \Sigma_t \int_0^{\infty} \psi(\bm x, \bm \Omega,  
        s) \, ds = \Sigma_t \psi_c(\bm x, \bm \Omega) \,,
        \end{align}
and thus by Eq.\ (\ref{4.11b}), 
   \begin{align}
    \hat F(\bm x) = \Sigma_t \int_{4\pi} \psi_c(\bm x, \bm \Omega' ) \, d\Omega'
        = \Sigma_t \phi_c(\bm x) \,. 
        \end{align}
Using Eq.\ (\ref{6.7}) and the previous result in Eqs.\ (\ref{4.12b}) and (\ref{4.12c}), we obtain:
   \begin{subequations} \label{6.10}
   \begin{equation}
      \psi_c(\bm x, \bm \Omega) = \frac{1}{4\pi} \int_0^{\infty} \bigl[ \Sigma_s \phi_c(\bm x - s \bm \Omega) + Q(\bm x - s \bm \Omega) \bigr] e^{- \Sigma_t s} ds \,,
   \label{6.10a}
   \end{equation}    
   and
   \begin{equation}
      \phi_c (\bm x) = \iiint \left[ \Sigma_s \phi_c (\bm x') + Q(\bm x') \right]  
      \frac{e^{- \Sigma_t |\bm x - \bm x' |}}{4 \pi | \bm x - \bm x' |^2} \, dV' \,.
   \label{6.10b}
   \end{equation}
  \end{subequations}
Equation (\ref{6.10b}) is the classic integral transport equation for the scalar flux $\phi_c(\bm x)$, and Eq.\ (\ref{6.10a}) is the classic expression for the angular flux $\psi_c(\bm x, \bm \Omega)$ in terms of $\phi_c(\bm x)$, for the case of isotropic scattering.

For the theory involving the asymptotic diffusion limit, if Eq.\ (\ref{6.1}) holds, then we can
use Eqs.\ (\ref{5.34}) and (\ref{6.6}) to reduce Eq.\ (\ref{5.23})
to the classic diffusion expression: 
   \begin{equation}
      - \frac{1}{3 \Sigma_t(1-c\overline\mu_0)} \nabla^2 \Phi^{(0)}(\bm x)
         + \Sigma_t (1-c) \Phi^{(0)}(\bm x) = Q(\bm x) \,.
   \label{6.11}
   \end{equation} 

In short, we have shown that when Eq.\ (\ref{6.1}) holds, the new generalized theory reduces to the classic transport theory, as it must.

%%%%%%%%%%%%%%%%%%%%%%%%%%%%%%%%%%%%%%%%%%%%%%%%%

\vspace{20pt}
\noindent
{\bf 7. DISCUSSION}
\setcounter{section}{7}
\setcounter{equation}{0} 
\vspace{10pt}

We have generalized the theory introduced in \cite{larsen_07}, in which the true non-exponential probability distribution function for the distance-to-collision is replaced by its ensemble average. This ensemble-averaged probability distribution function is used at all points to determine how far particles travel between collisions. The original aspect of the present model is in allowing the cross sections of the homogenized system to be functions of both angle $\bm\Omega$ and distance-to-collision $s$; that is, we assume that the locations of the scattering centers are correlated {\em and} depend upon direction.

With this method, we obtain a new generalized Boltzmann equation that preserves all relevant asymptotic limits and reduces to the classic Boltzmann equation in the case of a homogeneous system. The disadvantage is that there is no simple way to obtain expressions for these $s$-dependent cross sections; they need to be numerically estimated. Nevertheless, we use this result to develop a generalized diffusion equation, which does not depend upon the variable $s$, but rather on its mean and mean-squared values $s_{\bm\Omega}$ and $s_{\bm\Omega}^2$. The diffusion approximation also preserves all relevant limits and reduces to classic diffusion in homogeneous systems. Moreover, it yields anisotropic diffusion coefficients when the locations of the scattering centers depend upon direction.

The present theory requires more information about a random system than
the atomic mix method; if a random system is diffusive, then only $s_{\bm\Omega}$ and $s_{\bm\Omega}^2$ need to be estimated (for isotropic diffusion, they will not depend on $\bm\Omega$). This extra information is microscopic in nature; it is not a closure relation, as in the Levermore-Pomraning method. Our theory uses this microscopic data in a generalized Boltzmann equation or in a generalized diffusion equation to determine approximate mean macroscopic quantities. In Part 2 of this paper we numerically show that for problems of the pebble bed kind, this new approach more accurately predicts the anisotropic behavior of the systems than the standard GLBE and the classic approaches currently in use.

In the future, we intend to extend the present work to problems with inhomogeneous statistics, as well as for energy-dependent systems. Also, having developed a theory for the mean flux of particles, the next logical step is to try to develop a similar theory that is capable of calculating variance; this is another of our goals.

%%%%%%%%%%%%%%%%%%%%%%%%%%%%%%%%%%%%%%%%%%%%%%%%%

\vspace{20pt}
\noindent
{\bf ACKNOWLEDGMENT}
\vspace{10pt}

Richard Vasques would like to thank CAPES - Coordena\c{c}\~ao de Aperfei\c{c}oamento de Pessoal de N\'ivel Superior and the Fulbright Program for financial and administrative support.

%%%%%%%%%%%%%%%%%%%%%%%%%%%%%%%%%%%%%%%%%%%%%%%%%

\vspace{14pt}
\setlength{\baselineskip}{14pt}
\renewcommand{\section}[2]{{\bf \vspace{10pt} \noindent REFERENCES\\}}

\end{document}